\documentclass[preprint,12pt]{elsarticle}

\usepackage{xcolor}
\usepackage{graphicx}
\usepackage{dcolumn}
\usepackage{bm}

\begin{document}


\title{Development of self-modulation as a function of plasma length}

\author{Arthur Clairembaud$^{a,b}$} 

\affiliation{organization={Max Planck Institute for Physics},
            city={Garching},
            state={Bavaria},
            country={Germany}}
\affiliation{organization={Technical University of Munich},
            city={Garching},
            state={Bavaria},
            country={Germany}}
\author{Marlene Turner$^{c}$} 
\affiliation{organization={CERN},
            city={Geneva},
            country={Switzerland}}
\author{Patric Muggli$^{a}$} 

\date{\today}

\begin{abstract}
We use numerical simulations to determine whether the saturation length of the self-modulation (SM) instability of a long proton bunch in plasma could be determined by measuring the radius of the bunch halo SM produces. %
Results show that defocused protons acquire their maximum transverse momentum and exit the wakefields at a distance approximately equal to the saturation length of the wakefields. %
This suggests that measuring the radius of the halo as a function of plasma length in the AWAKE experiment would yield a very good estimate for the saturation length of SM. %
\end{abstract}

\maketitle
\section{Introduction}

AWAKE aims at developing a plasma wakefield accelerator (PWFA~\cite{chen}) driven by an energetic, but long proton bunch to accelerate an externally-injected electron bunch to high energies (50-100\,GeV) for application to particle physics~\cite{matthew}. %

To reach an accelerating field amplitude of $\approx$1\,GV/m, the long bunch must experience self-modulation (SM)~\cite{kumar}. %
The accelerator would then consist of two plasmas~\cite{muggli,edda}: one for SM, and one for acceleration. %
The SM process grows from seed wakefields~\cite{fabian,livio}, and amplitudes suitable for large acceleration are reached only after the process saturates~\cite{pukhov,schroeder}. %
It is thus essential to determine the saturation length of the SM process to choose the length of the first plasma, the self-modulator. %
The saturation length is also a fundamental parameter of the SM process.%

We have deduced from experimental and simulation results that the SM process saturates between 3 and 5\,m of propagation of the proton bunch in the plasma~\cite{marlene,marleneprab}. %
One of the measurements used to reach this conclusion is the radius of the halo formed around the core of the bunch by defocused particles, as measured at a screen located downstream from the exit of the plasma. %
To lowest order, this halo formation is the consequence of the growth of SM and of the formation of the microbunch train (purely transverse process), and thus would stop at saturation of the SM process. %
Indeed, the microbunch train is then fully formed, and particles remaining in the wakefields remain so, and can in principle drive wakefields over a long distance. %
One can thus assume that halo particles with the largest perpendicular momentum $p_{\perp, max}$, determine the radius of the halo $r_{h}$, and leave the wakefields near their saturation length $L_{sat}$.

In this paper we use numerical simulation results to determine whether the dependencies of the radius of the halo versus plasma length, $r_h(L_p)$, can be used to determine $L_{sat}$. %
The expectation is thus that $r_{h}$ would increase when increasing $L_{p}$ and saturate for $L_p\cong L_{sat}$. %
For $L_p > L_{sat}$, $r_{h}$ would remain essentially constant. %

The results we present show that measuring the length of plasma at which $r_{h}$ saturates gives a very good estimate for the saturation length of the wakefields. %
They also confirm that for all values of $L_p\ge2$\,m, particles reaching the largest radius at the screen, i.e., acquiring the largest transverse momentum, experience first a focusing force, travel towards the bunch axis, cross it, and then experience a defocusing force and leave the wakefields, as was pointed out before~\cite{TURNER2018123}. %

These studies are motivated by the ability to vary $L_{p}$ in experiments, recently afforded by the installation screens that can be inserted every meter along the new plasma source~\cite{muggliaac22}. %
These screens can block the ionizing laser pulse, and thus determine the length of the plasma. %
The length can be varied in steps of 1\,m, giving $L_p$= 0.5, 1.5, ..., 9.5\,m, and 10.3\,m without screen. %
\section{Numerical simulation results}

\begin{figure}[htbp]
    \centering
    \includegraphics[width=1\linewidth]{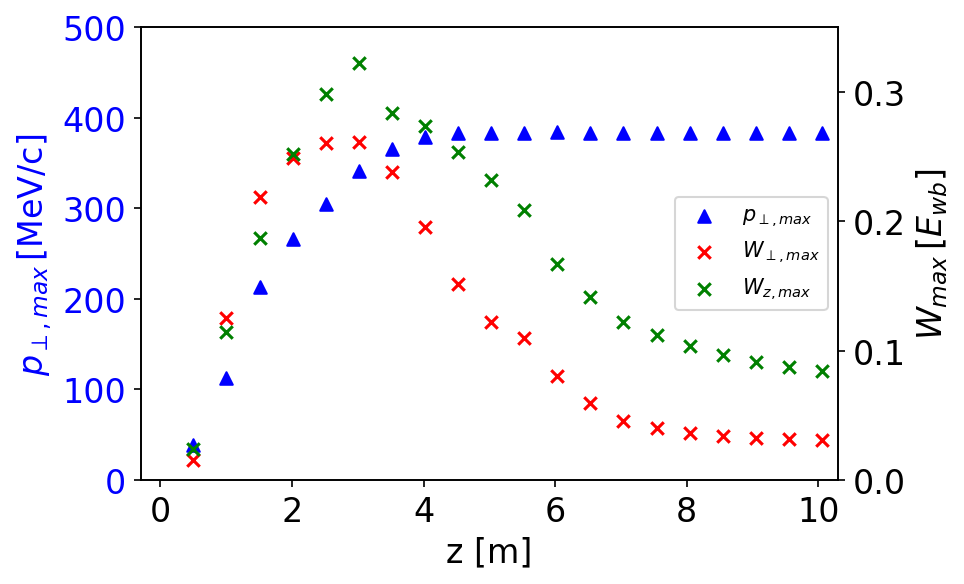}
    \caption{Green symbols: maximum longitudinal wakefield amplitude $W_{z,max}$ along the bunch and at $r=0$. %
               Red symbols: maximum transverse wakefield amplitude $W_{\perp,max}$ along the bunch and at $r=\sigma_{r0}$. %
               $W_{\perp,max}$, and $W_{z,max}$ in units of the cold wavebreaking field $E_{wb}=m_ec\omega_{pe}/e\cong2.5$\,GV/m. %
               Blue symbols: minimum value of the $1\%$ protons with the largest (positive) transverse momentum, $p_{\perp, max}$.}
    \label{fig:pr_max_and_W_r}
\end{figure}

Simulations were performed with the quasi-static, 2D3v code LCODE~\cite{LCODE}. %
Simulation parameters are given in Table~\ref{tablesim}. %
\begin{table}
    \centering
    \begin{tabular}{ccc}
         \hline
         Parameter                &  Units           & Value\\
         \hline
         Radial window size         & $c/\omega_{pe}$  & 8.0\\
         Radial plasma size         & $c/\omega_{pe}$  & 4.98\\
         Longitudinal window size   & $c/\omega_{pe}$  & 750\\
         Spatial resolution         & $c/\omega_{pe}$  & 0.0125\\
         Number of time steps       & $c/\omega_{pe}$  & 200\\
         Plasma particles per cell  &                  & 100\\
         Beam particles per slice   &                  & 200\\
         \hline
    \end{tabular}
    \caption{Numerical simulation parameters. %
    The proton bunch is initiated with $\cos^2$ longitudinal, and Gaussian transverse profiles.}
    \label{tablesim}
\end{table}
All simulations use the AWAKE baseline parameters given in Table~\ref{tablephys}. %
\begin{table}
    \centering
    \begin{tabular}{cccc}
         \hline
         Parameter& Variable&  Units& Value\\
         \hline
         Number of protons in the full bunch & $N_{p}$      &    -         &    $3\times10^{11}$\\
         RMS radius at the plasma entrance& $\sigma_{r0}$    &  \textmu m   & 165\\
         RMS duration& $\sigma_{t}$     &  ps          & 170   \\
         Normalized emittance& $\epsilon_N$     &  mm-mrad     & 2.2\\
         Lorentz factor& $\gamma$         &    -         & 427\\
         RIF timing & $t_{RIF}$$^{*}$ &     ps       &  0 \\
         Plasma electron density& $n_{pe}$         &   cm$^{-3}$   &  $7\times10^{14}$\\
         Angular plasma electron frequency& $\omega_{pe}$         &   rad/s   &  $1.49\times10^{12}$\\
         \hline
    \end{tabular}
    \caption{Physical parameters used for the numerical simulations. All parameters describe the proton bunch, besides $n_{pe}$ and $\omega_{pe}$, which describe the plasma. %
    $^{*}t_{RIF}$=0 means that in an experiment the plasma (created by a relativistic ionization front (RIF)) starts at the peak of the proton bunch. %
    In numerical simulations, only the corresponding second half of the proton bunch is simulated. %
    }
    \label{tablephys}
\end{table}
To determine the saturation length of SM, we look at the evolution of the maximum wakefields $W_{max}$ along the plasma ($z$). %
For a neutral plasma with constant density, the wakefields driven by the proton bunch are expected to first grow, saturate, and eventually decay along the plasma~\cite{kumar,pukhov,schroeder}. %
Figure~\ref{fig:pr_max_and_W_r} shows that, with the parameters of Table~\ref{tablephys}, this evolution occurs over 10\,m of plasma. %
The transverse ($W_{\perp,max}$, red crosses) and longitudinal ($W_{z,max}$, green crosses) wakefields are evaluated at their maximum along the bunch (which can be at a different positions ($\xi = z - v_b t$, where $v_b$ is the bunch velocity) for different $z$), and at radial positions $r=\sigma_{r0}$ and $r=0$, respectively. %
The growth of the wakefields ($z<$3\,m) occurs as the long, continuous, proton bunch self-modulates and the microbunch train, that resonantly excites wakefields, forms. %
This process leads to the saturation of SM ($z \cong$3\,m) 
when the microbunch train is fully formed, and wakefields reach their largest amplitude ($W_{\perp,max}\cong$0.25 and $W_{z,max}\cong$0.30\,$E_{wb}$). %

After saturation 
($z>$3\,m), the evolution of SM, i.e. of the wakefields and microbunch train, continues; the phase of the wakefields with respect to the train continues to shift backwards, and particles can find themselves in the defocusing phase of the wakefields. %
As this process occurs, most particles are eventually defocused, and only few particles remain on axis. %
Consequently, the wakefield amplitude driven by the remaining train decreases. We note that the presence of a step in the plasma density during the growth of SM can counter this decrease~\cite{lotovstep}.%

Because SM is a transverse process, the formation of the microbunch train leads to the formation of a halo of defocused protons, surrounding the train. %
Throughout the development of SM, these defocused protons acquire a transverse momentum ($\vec{p_\perp}=\vec{p_r}+\vec{p_\theta}$
) that is proportional to the wakefield amplitude they experience: %
\begin{equation}
    \vec{p}_{\perp}(L) = \frac{q}{c} \int_{0}^{L} \vec{W}_\perp(z,\xi,r)dz + \vec{p}_{\perp0}(z=0,r),
    \label{eq:pr_eq}
\end{equation}
where $L$ is the distance over which they experience the fields $W_{\perp}(z,\xi,r) = E_r(z,\xi,r) - v_p B_\theta (z,\xi,r)$, and $p_{\perp0}$ their initial momentum (from the bunch emittance). %
We observe that the transverse momentum of the most defocused protons (i.e., with $p_\perp$ away from the axis) of the bunch $p_{\perp,max}$ (blue triangles, Fig.~\ref{fig:pr_max_and_W_r}), first increases as SM develops and the fields increase ($z<$3 to 4\,m), saturates approximately where the amplitude of the fields also saturates ($z \cong 4\,$m), and remains constant after that ($z>$4\,m). %

To understand this $p_{\perp, max}(z)$ dependency, we look at the transverse momentum distribution $N_b(p_{\perp})$ at different locations along the plasma.%
\begin{figure}[htbp]
    \centering
    \includegraphics[width=1\linewidth]{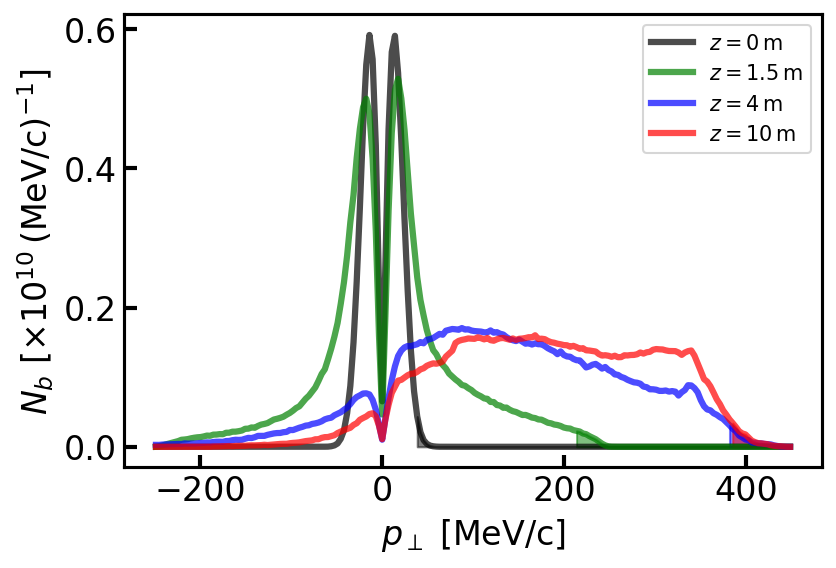}
    \caption{Transverse momentum distribution $N_b(p_\perp)$ of the proton bunch at different locations along the plasma. %
    Curves: number of protons per bin, bin size: $3.5\,$MeV/c. %
    Shaded areas under the curves: protons with a larger transverse momentum than the minimum $p_\perp$ of the $1\%$ most defocused ones ($p_{\perp,max}>0$). %
    Distribution at the plasma entrance $N_b(p_\perp)(z=0)$ (black line) divided by a factor 2 to improve the readability of the figure. %
    }
    \label{fig:p_perp_distribution}
\end{figure}
Figure~\ref{fig:p_perp_distribution} shows that between $z$=0 (black curve) and $z$=4\,m (blue curve), protons acquire transverse momentum (positive and negative, i.e., away and towards the axis), and the value of $p_{\perp,max}$ increases (shaded area under the curves, $\cong$220\,MeV/c, at $z$=1.5\,m). %

After saturation ($z>$4\,m) however, some protons continue to acquire transverse momentum; there are more protons with a $p_\perp$ value  between 100 
and 380\,MeV/c at $z$=10\,m (red curve) than at $z$=4\,m (blue curve), but $p_{\perp,max}$ does not 
increase ($\cong$400\,MeV/c, at $z$=4 and 10\,m). %
This suggests that 
particles with the largest transverse momentum have exited the wakefields transversely around the saturation point of SM, and are thus not affected anymore by the additional propagation. 
It also suggests that for $z>$4\,m, 
the particles that remain in the wakefields can gain transverse momentum but do not reach values larger than $p_{\perp,max}(z=4\,m)$. %

To confirm this, we track the $p_{\perp}$ evolution of protons that start drifting ballistically at different ranges along the plasma ($L_d$). %
\begin{figure}[htbp]
    \centering
    \includegraphics[width=1\linewidth]{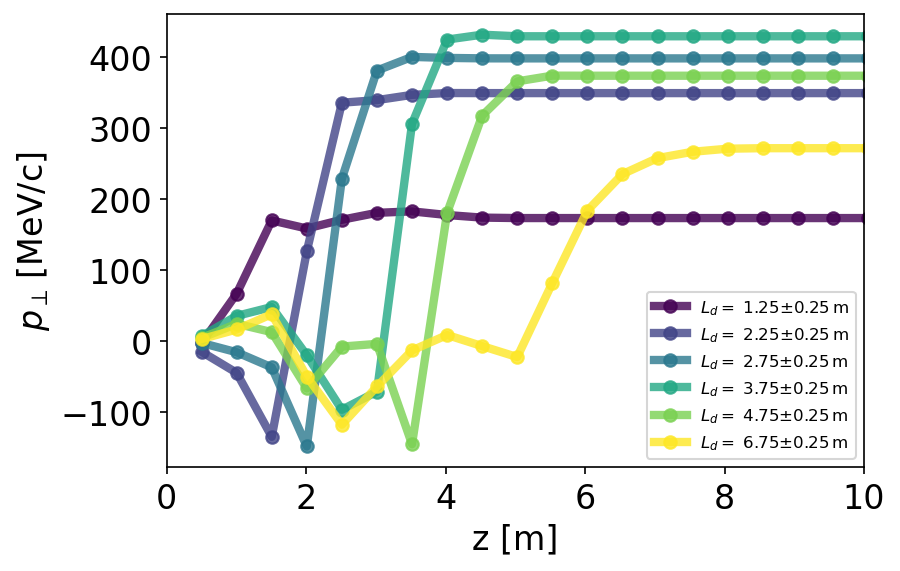}
    \caption{Transverse momentum ($p_\perp$) of the protons that exit over different ranges $L_d$ along the plasma. %
    Each curve is the average $p_\perp$ of the 1000 particles that exit within $L_d$ with the largest $p_\perp$. %
    $L_d$ is defined as the location after which the $p_\perp$ of protons does not change by more than $10\%$ (from $z=L_d$ to $z=10\,m$), i.e., they are traveling ballistically, outside of the wakefields, from that point on. %
    Negative values of $p_\perp$ mean particles are traveling toward the axis. %
    }
    \label{fig:p_perp_vs_z_for_particles}
\end{figure}
Figure~\ref{fig:p_perp_vs_z_for_particles} shows that the particles that exit the wakefields over $L_d=(1.25\pm0.25)\,$m (purple symbols), have a smaller $p_\perp$ value ($\cong 170\,$MeV/c) than $p_{\perp,max}(z=1.5)$ ($\cong 220$\,MeV/c, Fig.\ref{fig:p_perp_distribution}). %
This shows that the particles with the largest transverse momentum at $z$=1.5\,m are still in, and experiencing the wakefields. 
However, we observe that, around saturation, $L_d$=(3.75$\pm$0.25)\,m (dark green symbols)
, strongly defocused protons with $p_\perp \cong p_{\perp,max}$ exit the wakefields, i.e. their momentum is constant for $z>$4\,m. 
After saturation, e.g. 
$L_d$=(4.75$\pm$0.25)\,m (light green symbols) protons that exit the wakefields do so with a smaller transverse momentum than the ones that exited the wakefields around saturation, and the transverse momentum of the most defocused protons of the bunch is thus unchanged along the plasma: $p_{\perp,max}(z>4)$ is constant. %
This is true for all $L_d>L_{sat}$ (not shown). %
These two observations confirm that the particles that have the largest momentum do indeed leave the wakefields around $4$\,m, and that those still in the wakefields beyond this point do not acquire a larger transverse momentum.%

Additionally, we observe that the protons that exit the wakefields transversely with the largest $p_\perp$ ($L_d$=(3.75$\pm$0.25)\,m) acquire their $p_\perp$ between $z\cong$2 and 4\,m, i.e. around the saturation point of the fields $L_{sat}\cong$3\,m. %
That is because $p_\perp$ is proportional to the integral of the fields, and thus, when 
accumulating over a given distance, reaches its maximum value right after saturation of the fields. %
This explains why $p_{\perp,max}$ saturates slightly later ($z\cong$4\,m) than the wakefields ($z\cong$3\,m).%

We showed above that the position at which $p_{\perp,max}$ saturates as a function of $z$, 
is close to the position at which the value of the wakefields, and thus SM saturates. %

Evaluating $p_{\perp,max}$ at different locations along the plasma is equivalent to evaluating $p_{\perp,max}$ at the plasma exit for different plasma lengths ($L_p$). %


\begin{figure}[htbp]
    \centering
    \includegraphics[width=1\linewidth]{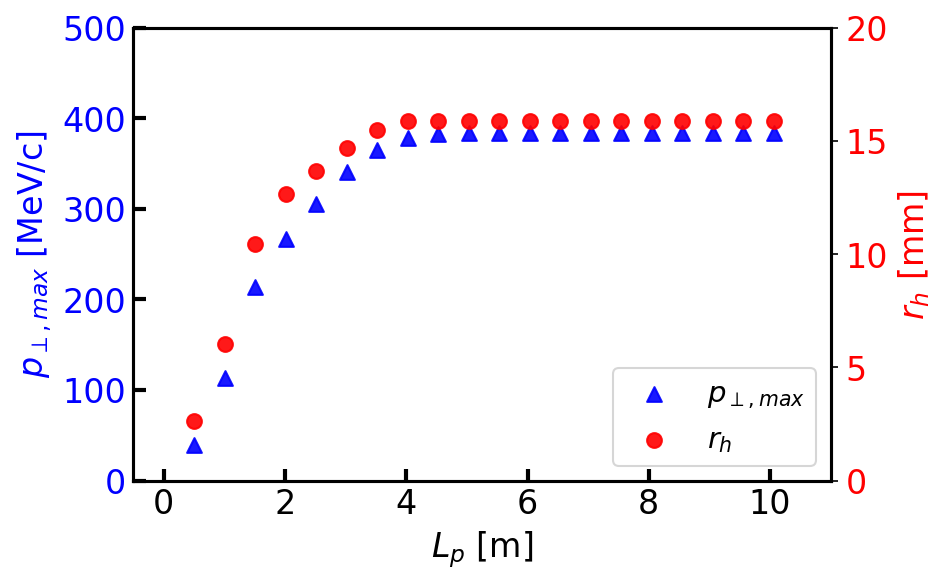}
    \caption{Blue symbols: minimum value of the $1\%$ protons with the largest (positive) transverse momentum, $p_{\perp, max}$. %
    Red symbols: radius of the halo after ballistic propagation of the protons from the plasma exit $z=L_p$ to a fixed screen $z=L_s=20\,$m from the plasma entrance. %
    }
    \label{fig:p_perp_vs_size}
\end{figure}

In an experiment however, $p_{\perp,max}$ cannot be measured directly. In order to relate $p_{\perp,max}$ to a parameter that can be measured, we propagate protons ballistically from the location at which they exit the wakefields ($z=L_d \le L_p$) to a fixed location after the plasma entrance ($z = L_s=$20\,m, location of a screen from the plasma entrance). %
The radial position reached by each proton at the screen is thus $\vec{r}_s = \vec{r}_0 + (\vec{p}_{\perp}/\gamma m_{p})(L_s-L_d)/v_b$, where $r_0$ is the radial position at which protons exit the wakefields (typically a few cold plasma skin-depths $c/\omega_{pe}\sim$200\,\textmu m), and $m_p$ is the mass of the protons. %
We measure the transverse extent of the proton bunch distribution at the screen (or halo radius), $r_h = \sqrt{(r_0 + \frac{p_r}{\gamma m_{p}}(L_s-L_d)/v_b)^2 + (\frac{p_\theta}{\gamma m_{p}})^2((L_s-L_d)/v_b)^2}$, as it is done in \cite{TURNER2018123,marlene}.%

Figure~\ref{fig:p_perp_vs_size} shows that for any $L_p$, $r_h \gg r_0$, i.e. $r_h \cong (p_{\perp}/\gamma m_{p})(L_s-L_d)/v_b$. We note that when $L_p$ increases from $0.5\,$ to $4\,$m, the momentum acquired by defocused protons increases but the propagation distance to the screen $(L_s-L_d)$ decreases. In other words, a proton that exits the wakefields with a given $p_\perp$ early along the plasma could, in principle, reach a larger $r_s$ at the screen than a proton that leaves later with a larger $p_\perp$. However, we observe that, with $L_s =$20\,m, this phenomenon does not occur, as the dependencies of $p_{\perp,max}$ versus $L_p$ dominate the ones of $(L_s-L_d)$. Finally, Figure~\ref{fig:p_perp_vs_size} confirms that the radius of the halo $r_{h}$ (red circles) and $p_{\perp,max}$ (blue triangles) show the same dependency versus $L_{p}$. %

Since $r_h$ saturates at the same position along the plasma as $p_{\perp,max}$ (Fig.~\ref{fig:p_perp_vs_size}), and the amplitude of the wakefields saturates at approximately the same location as $p_{\perp,max}$ (Fig.\ref{fig:pr_max_and_W_r}), the amplitude of the wakefields saturates at approximately the same location as $r_h$. We can therefore conclude that measuring $r_h(L_p)$ is equivalent to measuring $p_{\perp,max}(z)$, and that the saturation length of SM can be determined experimentally by measuring the position at which the halo radius saturates as a function of $L_p$.%

While we present results for only one set of parameters, initial simulation results show that this method allows, in principle, to measure the variations of the saturation length of the SM process for other sets of parameters ($n_{pe}$, $N_p$, $t_{RIF}$, ...), as well as the effect of a plasma density step on this length. %

\section{Conclusions}

In this paper, we show that we can measure the effect of transverse wakefields on protons by measuring, for various lengths of plasma, the %
transverse extent of the proton bunch after %
its propagation to a screen. 
We expect this transverse extent to first increase as a function of plasma length, and then %
to saturate. %
We show that it saturates because the most defocused protons exit the wakefields transversely around the saturation point of SM. Protons that exit the wakefields beyond this point do so with a smaller transverse momentum, and thus reach only a smaller radius. %
Finally, we show that the plasma length for which the radius of the bunch saturates is approximately that where the amplitude of the wakefields %
also saturates. 
This measurement thus provides a method to determine the saturation length of the self-modulation process in an experiment.



\bibliographystyle{elsarticle-num-names}
\bibliography{refs} 

\end{document}